\documentclass[a4paper,twocolumn,english]{revtex4}
\usepackage{babel,amsmath,amssymb,dcolumn}
\usepackage[dvips]{graphics}
\usepackage{graphicx}
\usepackage{setspace}[2]
\begin{document}

\title{Solutions for Klein-Gordon equation in Randall-Sundrum-Kerr scenario}

\author{ J\'eferson de Oliveira,
  Carlos Eduardo Pellicer de
  Oliveira}

\email{carlosep@fma.if.usp.br}

\affiliation{Instituto de F\'{\i}sica, Universidade de S\~{a}o Paulo \\
C.P. 66318, 05315-970, S\~{a}o Paulo-SP, Brazil}


%
%

\begin{abstract}
We study the scalar perturbations of rotating black holes in
framework of extra dimensions type Randall-Sundrum(RS).
\end{abstract}

\maketitle

We study the scalar perturbations of a rotating black hole, in a framework of extra dimensions described by a Randall-Sundrum model\cite{rs1}.
The extraction of energy from rotating black holes is possible due to
the well-known effect of superradiance\cite{Starob1,Starob2,Zeld1}: the reflected wave has an
amplitude larger than the incoming wave, and thereby is amplified. We
consider the process of superradiance for this background.
The conditions for superradiance and the reflection coefficients are
found for scalar field perturbations.

We analyze the scalar perturbations of the metric
\begin{equation}
ds^{2}=l^{2}/y^{2}\left[ds^{2}_{Kerr}+dy^{2}\right],
\end{equation}         
in which $ds^{2}_{Kerr}$ is the Kerr metric in the Boyer-Lindquist
coordiantes $(t,r,\theta,\phi)$.

The equation for scalar field dynamics $\Phi(X^{\mu},y)$, ($X^{\mu}=t,
r, \theta, \phi$) with mass $\mu^{2}$ is the Klein-Gordon equation
\begin{equation}\label{k9}
\frac{1}{\sqrt{-g}}\frac{\partial}{\partial
x^{A}}\left(g^{AB}\sqrt{-g}\frac{\partial\Phi}{\partial
x^{B}}\right)+\mu^{2}\Phi=0.
\end{equation}
Due to the axial symmetry and stationarity of the background metric we
can expand the scalar field in the appropriate harmonics which are
defined by the symetry group of motion of the metric,   
$
\Phi(X^{\mu},y)=R(r)S(\theta)\Omega(y)e^{im\phi}e^{-i\omega t}.
$
After some algebra we obtain two equations, thereby decoupling the
bulk evolution equation from the genuine brane fields. The angular and radial variables can be easily
separated by the usual methods \cite{kokkotas1}. In the end, we obtain three equations with separated
variables to radial, angular, and bulk coordinates,  
\[
\frac{d}{dr}\left(\Delta\frac{dR(r)}{dr}\right)+\frac{R(r)}{\Delta}[(r^{2}+a^{2})^{2}\omega^{2}-4Mram\omega
+
\]
\begin{equation}\label{k15}
+(ma)^{2}-\Delta\left((a\omega)^{2}+Q^{2}r^{2} + P \right)]=0,
\end{equation}
\[
\frac{1}{\sin\theta}\frac{d}{d\theta}\left(\sin\theta\frac{dS(\theta)}{d\theta}\right)+S(\theta)\left[
(\omega^{2}-Q^{2})a^{2}\cos^{2}\theta\right]
\]
\begin{equation}
\label{k17}
-S(\theta)\left[\frac{m^{2}}{\sin^{2}\theta}+P\right]=0,
\end{equation}
\begin{equation}
\label{b18}
\frac{d}{dy}\left(f(y)^{3/2}\frac{d\Omega(y)}{dy}\right)+Q^{2}\Omega(y)f(y)^{3/2}+\mu^{2}f(y)=0.
\end{equation}      
Let us make the usual replacements in order to reduce Eq.(\ref{k15}) to the standard wave-like type
\begin{eqnarray}\label{k16}
\frac{d^{2}\Psi(r)}{dr_{*}^{2}}+\Psi(r)\left[\omega^{2}-V(r,\omega)\right]=0,
\end{eqnarray}
with the potential for massless scalar perturbation defined by
\[
V(r,\omega)=\frac{4Mra(m\omega)-(ma)^{2}+\Delta((a\omega)^{2}
+Q^{2}r^{2}+P)}{(r^{2}+a^{2})^{2}} +
\]
\begin{equation}
\frac{\Delta(3r^{2}-4Mr+a^{2})}{(r^{2}+a^{2})^{3}}-\frac{3\Delta^{2}r^{2}}
{(r^{2}+a^{2})^{4}}.
\end{equation}

The branes are situated at $z=0$ and $z=d$ (or at $y=l$
and $y=l e^{d/l}$ respectively). 
Therefore, the solution for equation (\ref{b18}) is
\begin{equation}
\label{k18}
\Omega(y)=y^{2}\left[AJ_{\sqrt{4-k^{2}}}(Qy)+BY_{\sqrt{4-k^{2}}}(Qy)\right],
\end{equation}
in which $A$, $B$ are constants, $J_{\sqrt{4-k^{2}}}(Qy)$,
$Y_{\sqrt{4-k^{2}}}(Qy)$ are Bessel's functions and $k^{2}=\mu^{2} l^{2}$.

In the two-brane world model, the boundary conditions for the
perturbations comes from the Israel junction conditions \cite{israel}
which imply the constraint for the massless scalar field
\begin{equation}\label{i2}
Y_{1}(Q_{n})J_{1}(x_{n})=Y_{1}(x_{n})J_{1}(Q_{n}l), \quad
Q_{n}=\frac{x_{n}}{l}e^{l/d}.
\end{equation}
%
%

%
%
%
We note that at spatial infinity the ``effective
potential'' $\omega^2 - V$  has the asymptotic form
\begin{equation}\label{s1}
\omega^2 - V \rightarrow \omega^2 - Q^{2}, \quad  r^{*} \rightarrow +\infty,
\end{equation}
while at the event horizon the asymptotic form of the equation is  
\begin{equation}
\omega^2 - V \rightarrow (\omega - m \Omega)^2, \quad  r^{*} \rightarrow -\infty,
\end{equation}
in which $ \quad 
\Omega = \frac{a}{2 M r_{+}}, \quad r_{+} = M + \sqrt{M^2 - a^2}.
$
In equation (\ref{s1}), we see that when $Q^{2}$ is larger than $\omega^2$
the effective potential is clearly negative at sufficiently large
$r$ and the scalar field perturbations are unstable. Thus, we need to
be restricted by the case when $Q^{2} < \omega_{qn}^2$, where
$\omega_{qn}$ is the lowest (fundamental) quasinormal mode. In that case, the
asymptotic solutions of the wave equation (\ref{k16}) have the form:
\begin{equation}
\Psi \rightarrow B_{L m \omega} e^{- i (\omega - m \Omega) r^{*}}, \quad  r^{*} \rightarrow -\infty 
\end{equation}
\begin{equation}
\Psi \rightarrow e^{- i (\sqrt{\omega^2 - Q^{2}}) r^{*}} +  A_{L m \omega}
e^{i (\sqrt{\omega^2 - Q^{2}}) r^{*}}, 
\quad  r^{*} \rightarrow +\infty. 
\end{equation}
Here the requirement that the wave should have an ingoing group velocity 
at the event horizon is satisfied. The wave comes from infinity,
partially passes through the potential barrier reaching the
event horizon, the rest reflects back. From the constancy of the
Wronskian, i.e. from equality of the Wronskian at both asymptotics,  
we have: 
$ \quad   
1- |A_{L m \omega}|^2 - |B_{L m \omega}|^2 \frac{\omega - m \Omega}{\sqrt{\omega^2 - Q^{2}}} = 0. 
$
It means that, similar to the ordinary Kerr case,  $|A_{L m \omega}|
>1$, i.e. the amplitude of the reflected wave is larger than that  of
the incident wave if the following condition of superradiance takes place:
\begin{equation}    
m \Omega > \omega. 
\end{equation}
In order to find the reflection coefficient let us consider the near
region wave behavior, when $r - r_{+} \ll 1/\omega$. In this approximation 
Eq.(\ref{k15}) reads:  
\begin{equation}\label{s2}
\Delta\frac{d}{dr}\left(\Delta\frac{dR(r)}{dr}\right)
+[r_{+}^{4} (\omega - m \Omega)^{2} - L (L+ 1) \Delta] R(r)=0.
\end{equation}
The general solution of this equation is 
\[
R = A z ^{-i \chi} (1-z )^{L+1} F(a-c+1, b-c+1, 2-c, z)+ 
\]
\begin{equation}
B z ^{i \chi}
(1-z )^{L+1} F(a, b, c, z), 
\end{equation}
in which
\begin{equation}\label{s5}
\chi = (\omega - m \Omega) \frac{r_{+}^{2}} {r_{+}-r_{-}}, \quad
z= \frac{r - r_{+}}{r - r_{-}}.
\end{equation}
Following \cite{Starob1} we obtain the reflection coefficient $b / a$ 
\[
\frac{b} {a} = 2 i (\omega^{2}-Q^{2})^{L +1/2}  \chi \frac{(1-)^L}{2L
  +1} (\frac{L!}{(2 L
-1)!})^{2} 
\]
\begin{equation}\label{s4}
\times\frac{(r_{+}-r_{-})^{2 L+ 1}}{(2L)! (2 L +1)!} (k^2 + 4 \chi^2).
\end{equation}

By estimating the possible influence of the  RS model onto
superradiance let us find the eigenvalues $Q_{n}$  of  the bulk
equation (\ref{k18}). We can solve equation (\ref{i2})
numerically. The spectrum of egenvalues $Q_{n}$  does not depend 
on $d$, and depend on  $l$ very mildy, provided that
$l$ is small.

The first ten modes for $l=0.0001$ and $d=0.0005$ m. are shown in the Table I.
\begin{table}[ht] 
\caption{$Q_{n}$ eigenvalues} 
\begin{tabular}{|l|l||l|l|} 
\hline
$n$  & $Q_{n}$ & $n$ & $Q_{n}$  \\
\hline
 1 &   3.83171   &  6 & 19.6159  \\
\hline 
 2  &  7.01559  &  7 &22.7601 \\
\hline
 3 & 10.1735 & 8 & 25.9037  \\
\hline
4 &13.3237  & 9 &  29.0468\\
\hline 
5& 16.4706 & 10 &32.1897 \\
\hline
\end{tabular}
\end{table}
For black holes of one tenth of solar mass we definitely  avoid GL
instability \cite{gl}. In spite of this, the superradiant instability hinted by
\cite{cardoso1} appears, because this massless scalar perturbation si\-mulate a
massive parturbation in the $4-$dimensional Kerr black hole which is unstable. 


{\bf{Acknowledgments}}: We wish to acknowledge useful discussions with Roman Konoplya, Karlucio H. Castello-Branco and Prof. Elcio Abdalla. This work is supported by FAPESP. 

\end{document}